\documentclass[12pt]{article}

\usepackage{amsmath,amssymb,amscd,amsbsy,amsgen,amsopn,amstext,
amsxtra}

\usepackage[dvips]{graphicx}

\begin{document}

\begin{flushright}
\end{flushright}

\vskip 0.5 truecm

\begin{center}
{\Large{\bf Topological properties of geometric phases\footnote{Talk given at the International Workshop "Frontiers in Quantum Physics", Yukawa Institute for Theoretical Physics, Kyoto,
Japan, February 17-19, 2005 (to be published in the Proceedings).}}}
\end{center}
\vskip .5 truecm
\centerline{\bf Kazuo Fujikawa }
\vskip .4 truecm
\centerline {\it Institute of Quantum Science, College of 
Science and Technology}
\centerline {\it Nihon University, Chiyoda-ku, Tokyo 101-8308, 
Japan}
\vskip 0.5 truecm

\makeatletter
\@addtoreset{equation}{section}
\def\theequation{\thesection.\arabic{equation}}
\makeatother

\begin{abstract}
The level crossing problem and associated geometric terms are
neatly formulated by using the second quantization technique
 both in the operator and path integral formulations.
The analysis of geometric phases is then reduced to the familiar 
diagonalization of the Hamiltonian. If one diagonalizes the 
Hamiltonian in one specific limit, one recovers the conventional
formula for geometric phases. On the other hand, if one diagonalizes the geometric terms in the infinitesimal 
neighborhood of level crossing, the geometric phases become 
trivial (and thus no monopole singularity) for arbitrarily large
 but finite time interval $T$. The topological 
proof of the Longuet-Higgins' phase-change rule, for example, thus fails in the practical Born-Oppenheimer approximation where a large but finite ratio of
 two time scales is involved and $T$ is identified with the 
period of the slower system.
\end{abstract}

\section{ Berry's Phase}

We start with a brief account of the Berry's phase or 
geometric phase~\cite{pancharatnam}-\cite{bhandari}.
We study the Schr\"{o}dinger equation
\begin{eqnarray}
i\hbar\frac{\partial}{\partial t}\psi(t,\vec{x})
=\hat{H}(\frac{\hbar}{i}\frac{\partial}{\partial\vec{x}},
\vec{x},X(t))
\psi(t,\vec{x}) 
\end{eqnarray}
where the Hamiltonian depends on a set of slowly varying 
external parameters $X(t)=(X_{1}(t), X_{2}(t), X_{3}(t), ...)$.
By analyzing this simple Schr\"{o}dinger equation in the 
adiabatic approximation, one finds an extra phase factor in addition 
to the conventional dynamical phase. There is no mystery about 
the phase factor, however, and all the information about the phase factor is included in the standard 
Schr\"{o}dinger equation and the evolution operator.   
 It is sometimes stated in the literature that there appears a magnetic monopole at the level crossing point, though the starting Hamiltonian 
$\hat{H}(\frac{\hbar}{i}\frac{\partial}{\partial\vec{x}},\vec{x},X(t))$ 
contains no obvious singularity. We are  
going to explain that there is no magnetic monopole at the 
level crossing point. To discuss these issues, we need a suitable
formulation which makes the analysis simple and transparent.

\section{ Second quantized formulation of level crossing}

We start with the generic hermitian Hamiltonian 
\begin{eqnarray}
\hat{H}=\hat{H}(\hat{\vec{p}},\hat{\vec{x}},X(t))
\end{eqnarray}
in a slowly varying background variable 
$X(t)=(X_{1}(t),X_{2}(t),...)$.
The path integral for this theory for the time interval
$0\leq t\leq T$ in the second quantized 
formulation is given by 
\begin{eqnarray}
Z&=&\int{\cal D}\psi^{\star}{\cal D}\psi
\exp\{\frac{i}{\hbar}\int_{0}^{T}dtd^{3}x[
\psi^{\star}(t,\vec{x})i\hbar\frac{\partial}{\partial t}
\psi(t,\vec{x})\nonumber\\
&&-\psi^{\star}(t,\vec{x})
\hat{H}(\hat{\vec{p}},\hat{\vec{x}},X(t))\psi(t,\vec{x})] \}.
\end{eqnarray}
We next define a complete set of eigenfunctions
\begin{eqnarray}
&&\hat{H}(\hat{\vec{p}},\hat{\vec{x}},X(0))u_{n}(\vec{x},X(0))
=\lambda_{n}u_{n}(\vec{x},X(0)), \nonumber\\
&&\int d^{3}xu_{n}^{\star}(\vec{x},X(0))u_{m}(\vec{x},X(0))=
\delta_{nm},
\end{eqnarray}
and expand 
$\psi(t,\vec{x})=\sum_{n}a_{n}(t)u_{n}(\vec{x},X(0))$.
We then have  $
{\cal D}\psi^{\star}{\cal D}\psi=\prod_{n}{\cal D}a_{n}^{\star}
{\cal D}a_{n}$
and the path integral is written as~\cite{fujikawa, fujikawa2} 
\begin{eqnarray}
Z&=&\int \prod_{n}{\cal D}a_{n}^{\star}
{\cal D}a_{n}
\exp\{\frac{i}{\hbar}\int_{0}^{T}dt[
\sum_{n}a_{n}^{\star}(t)i\hbar\frac{\partial}{\partial t}
a_{n}(t)\nonumber\\
&&-\sum_{n,m}a_{n}^{\star}(t)E_{nm}(X(t))a_{m}(t)] \}
\end{eqnarray}
where 
\begin{eqnarray}
&&E_{nm}(X(t))=
\int d^{3}x u_{n}^{\star}(\vec{x},X(0))
\hat{H}(\hat{\vec{p}},\hat{\vec{x}},X(t))u_{m}(\vec{x},X(0)).
\end{eqnarray}

We next perform a (time-dependent) unitary transformation
\begin{eqnarray}
a_{n}=U(X(t))_{nm}b_{m}
\end{eqnarray}
where 
\begin{eqnarray}
U(X(t))_{nm}=\int d^{3}x u^{\star}_{n}(\vec{x},X(0))
v_{m}(\vec{x},X(t))
\end{eqnarray}
with the instantaneous eigenfunctions of the Hamiltonian
\begin{eqnarray}
&&\hat{H}(\hat{\vec{p}},\hat{\vec{x}},X(t))v_{n}(\vec{x},X(t))
={\cal E}_{n}(X(t))v_{n}(\vec{x},X(t)), \nonumber\\
&&\int d^{3}x v^{\star}_{n}(\vec{x},X(t))v_{m}(\vec{x},X(t))
=\delta_{n,m}.
\end{eqnarray}
We emphasize that $U(X(t))$ is a unit matrix both at $t=0$ and 
$t=T$ if $X(T)=X(0)$, and thus $\{a_{n}\}=\{b_{n}\}$ both at 
$t=0$ and $t=T$. It is convenient to take the time $T$ as the period of the slower variable $X(t)$.

We can thus re-write the path integral as 
\begin{eqnarray}
&&Z=\int \prod_{n}{\cal D}b_{n}^{\star}{\cal D}b_{n}
\exp\{\frac{i}{\hbar}\int_{0}^{T}dt[
\sum_{n}b_{n}^{\star}(t)i\hbar\frac{\partial}{\partial t}
b_{n}(t)\nonumber\\
&&+\sum_{n,m}b_{n}^{\star}(t)
\langle n|i\hbar\frac{\partial}{\partial t}|m\rangle
b_{m}(t)
-\sum_{n}b_{n}^{\star}(t){\cal E}_{n}(X(t))b_{n}(t)] \}
\end{eqnarray}
where the second term in the action stands for the term
commonly referred to as Berry's phase. The second 
term is defined by
\begin{eqnarray} 
(U(t)^{\dagger}i\hbar\frac{\partial}{\partial t}U(t))_{nm}
&=&\int d^{3}x v^{\star}_{n}(\vec{x},X(t))
i\hbar\frac{\partial}{\partial t}v_{m}(\vec{x},X(t))\nonumber\\
&\equiv& \langle n|i\hbar\frac{\partial}{\partial t}|m\rangle.
\end{eqnarray}
In the operator formulation of the second quantized theory,
we thus obtain the effective Hamiltonian (depending on Bose or 
Fermi statistics)
\begin{eqnarray}
\hat{H}_{eff}(t)&=&\sum_{n}b_{n}^{\dagger}(t)
{\cal E}_{n}(X(t))b_{n}(t)
-\sum_{n,m}b_{n}^{\dagger}(t)
\langle n|i\hbar\frac{\partial}{\partial t}|m\rangle
b_{m}(t)
\end{eqnarray}
with $[b_{n}(t), b^{\dagger}_{m}(t)]_{\mp}=\delta_{n,m}$.
Note that these formulas are exact. 

It is shown~\cite{fujikawa, fujikawa2} that    
\begin{eqnarray}
&&\langle n|T^{\star}\exp\{-(i/\hbar)\int_{0}^{T}dt
\hat{{\cal H}}_{eff}(t)\}
|n\rangle\nonumber\\
&&=\langle n(T)|T^{\star}\exp\{-(i/\hbar)\int_{0}^{T}dt \hat{H}
(\hat{\vec{p}},\hat{\vec{x}},X(t))\}|n(0)\rangle
\end{eqnarray}
where $T^{\star}$ stands for the time ordering operation.
In our picture, all the phase factors are included in  the 
Hamiltonian in the Schr\"{o}dinger picture
\begin{eqnarray}
\hat{{\cal H}}_{eff}(t)&=&\sum_{n}b_{n}^{\dagger}(0)
{\cal E}_{n}(X(t))b_{n}(0)
-\sum_{n,m}b_{n}^{\dagger}(0)
\langle n|i\hbar\frac{\partial}{\partial t}|m\rangle
b_{m}(0).
\end{eqnarray}

\section{ Level crossing}

We are mainly interested in the topological properties in the 
infinitesimal neighborhood of level crossing. We thus assume 
that the level crossing takes place only between the lowest two 
levels, and we consider the familiar idealized model with only 
the lowest two levels. This approximation is expected to be 
valid in the infinitesimal neighborhood of the specific level
crossing. The effective Hamiltonian to be analyzed 
in the path integral is then defined  by the $2\times 2$
matrix 
\begin{eqnarray}
 h(X(t))=\left(E_{nm}(X(t))\right)
\end{eqnarray}
in the notation of (2.4). If one assumes that the level crossing takes place at the origin of the parameter space $X(t)=0$, one needs to analyze the matrix
\begin{eqnarray}
h(X(t)) = \left(E_{nm}(0)\right) + 
\left(\frac{\partial}{\partial X_{k}}E_{nm}(0)\right) X_{k}(t)
\end{eqnarray}
 for sufficiently small $(X_{1}(1),X_{2}(1), ... )$.  After a 
suitable re-definition of the parameters by taking linear 
combinations of  $X_{k}(t)$, we write the matrix as
\begin{eqnarray}
h(X(t))
&&=\left(\begin{array}{cc}
            E(0)+y_{0}(t)&0\\
            0&E(0)+y_{0}(t)
            \end{array}\right)
+ g \sigma^{l}y_{l}(t)
\end{eqnarray}
where $\sigma^{l}$ stands for the Pauli matrices, and $g$ is a 
suitable (positive) coupling constant.
 
The above matrix is diagonalized in a standard manner 
\begin{eqnarray} 
h(X(t))v_{\pm}(y)=(E(0)+y_{0}(t) \pm g r)v_{\pm}(y)
\end{eqnarray}
where $r=\sqrt{y^{2}_{1}+y^{2}_{2}+y^{2}_{3}}$  and
\begin{eqnarray}
v_{+}(y)=\left(\begin{array}{c}
            \cos\frac{\theta}{2}e^{-i\varphi}\\
            \sin\frac{\theta}{2}
            \end{array}\right), \ \ \ \ \ 
v_{-}(y)=\left(\begin{array}{c}
            \sin\frac{\theta}{2}e^{-i\varphi}\\
            -\cos\frac{\theta}{2}
            \end{array}\right)
\end{eqnarray}
by using the polar coordinates, 
$y_{1}=r\sin\theta\cos\varphi,\ y_{2}=r\sin\theta\sin\varphi,
\ y_{3}=r\cos\theta$. Note that
$v_{\pm}(y(0))=v_{\pm}(y(T))$ if $y(0)=y(T)$ except for 
$(y_{1}, y_{2}, y_{3}) = (0,0,0)$, and $\theta=0\ {\rm or}\ \pi$.
If one defines
\begin{eqnarray} 
v^{\dagger}_{m}(y)i\frac{\partial}{\partial t}v_{n}(y)
=A_{mn}^{k}(y)\dot{y}_{k}
\end{eqnarray}
where $m$ and $n$ run over $\pm$,
we have
\begin{eqnarray}
A_{++}^{k}(y)\dot{y}_{k}
&=&\frac{(1+\cos\theta)}{2}\dot{\varphi},
\nonumber\\
A_{+-}^{k}(y)\dot{y}_{k}
&=&\frac{\sin\theta}{2}\dot{\varphi}+\frac{i}{2}\dot{\theta}
=(A_{-+}^{k}(y)\dot{y}_{k})^{\star}
,\nonumber\\
A_{--}^{k}(y)\dot{y}_{k}
&=&\frac{1-\cos\theta}{2}\dot{\varphi}.
\end{eqnarray}

The effective Hamiltonian  is then given by 
\begin{eqnarray}
\hat{H}_{eff}(t)&=&(E(0)+y_{0}(t) + g r(t))b^{\dagger}_{+}b_{+}
\nonumber\\
&&+(E(0)+y_{0}(t) - g r(t))b^{\dagger}_{-}b_{-}
\nonumber\\
&& -\hbar \sum_{m,n}b^{\dagger}_{m}A^{k}_{mn}(y)\dot{y}_{k}b_{n}
\end{eqnarray}
which is exact in the present two-level truncation.

In the conventional adiabatic approximation, one approximates
the effective Hamiltonian by
\begin{eqnarray}
\hat{H}_{eff}(t)&\simeq& (E(0)+y_{0}(t) + g r(t))
b^{\dagger}_{+}b_{+}\nonumber\\
&&+(E(0)+y_{0}(t) - g r(t))b^{\dagger}_{-}b_{-}\nonumber\\
&&-\hbar [b^{\dagger}_{+}A^{k}_{++}(y)\dot{y}_{k}b_{+}
+b^{\dagger}_{-}A^{k}_{--}(y)\dot{y}_{k}b_{-}]
\end{eqnarray}
which is valid for $Tg r(t)\gg \hbar\pi$, the magnitude of the 
geometric term.
The Hamiltonian for $b_{-}$, for example, is then eliminated by 
a ``gauge transformation''
\begin{eqnarray}
b_{-}(t)=
\exp\{-(i/\hbar)\int_{0}^{t}dt[
E(0)+y_{0}(t) - g r(t)
-\hbar A^{k}_{--}(y)\dot{y}_{k}] \} \tilde{b}_{-}(t)
\end{eqnarray}
in the path integral with the approximation (3.9), and the amplitude 
$\langle 0|\hat{\psi}(T)b^{\dagger}_{-}(0)|0\rangle$, which 
corresponds to the probability amplitude in the first 
quantization, is given by
\begin{eqnarray}
\langle 0|\hat{\psi}(T)b^{\dagger}_{-}(0)|0\rangle&=&\exp\{-\frac{i}{\hbar}\int_{0}^{T}dt[
E(0)+y_{0}(t) - g r(t) 
-\hbar A^{k}_{--}(y)\dot{y}_{k}] \}
\nonumber\\
&&\times v_{-}(y(T))
\langle 0|\tilde{b}_{-}(T)\tilde{b}^{\dagger}_{-}(0)|0\rangle
\end{eqnarray}
with $\langle 0|\tilde{b}_{-}(T)\tilde{b}^{\dagger}_{-}(0)
|0\rangle=1$.
For a $2\pi$ rotation in $\varphi$ with fixed $\theta$, for 
example, the geometric term (the last term on the exponential
of (3.11)) gives rise to the well-known factor~\cite{berry}
\begin{eqnarray}
\exp\{i\pi(1-\cos\theta) \}
\end{eqnarray}
without referring to the parallel transport and holonomy.

Another representation, which is useful to analyze the behavior
in the infinitesimal neighborhood of the level crossing point, is obtained by a further unitary transformation~\cite{fujikawa, fujikawa2}
\begin{eqnarray}
b_{m}=U(\theta(t))_{mn}c_{n}
\end{eqnarray}
 where $m,n$ run over $\pm$ with
\begin{eqnarray}
U(\theta(t))=\left(\begin{array}{cc}
            \cos\frac{\theta}{2}&-\sin\frac{\theta}{2}\\
            \sin\frac{\theta}{2}&\cos\frac{\theta}{2}
            \end{array}\right),
\end{eqnarray}
and the above effective Hamiltonian (3.8) is written as
\begin{eqnarray}
\hat{H}_{eff}(t)&&= (E(0)+y_{0}(t)+gr\cos\theta)
c^{\dagger}_{+}c_{+}\nonumber\\
&&+(E(0)+y_{0}(t)-gr\cos\theta)c^{\dagger}_{-}c_{-}\nonumber\\
&&-gr\sin\theta c^{\dagger}_{+}c_{-}
-gr\sin\theta c^{\dagger}_{-}c_{+}\nonumber\\
&&-\hbar\dot{\varphi} c^{\dagger}_{+}c_{+}.
\end{eqnarray}
In the above unitary transformation, an extra geometric
term $-U(\theta)^{\dagger}i\hbar\partial_{t}U(\theta)$ is 
induced by the kinetic term of the path integral 
representation. One can
confirm that this extra term precisely cancels the term 
containing $\dot{\theta}$ in $b^{\dagger}_{m}
A^{k}_{mn}(y)\dot{y}_{k}b_{n}$. 
We thus {\bf diagonalize} the geometric terms in this representation.

In the infinitesimal neighborhood of the level crossing point,
namely, for sufficiently close to the origin of the parameter 
space $(y_{1}(t), y_{2}(t), y_{3}(t) )$ but 
$(y_{1}(t), y_{2}(t), y_{3}(t))\neq (0,0,0)$, one has
\begin{eqnarray}
\hat{H}_{eff}(t)&\simeq& (E(0)+y_{0}(t))
c^{\dagger}_{+}c_{+}\nonumber\\
&&+(E(0)+y_{0}(t))c^{\dagger}_{-}c_{-}
-\hbar\dot{\varphi} c^{\dagger}_{+}c_{+}.
\end{eqnarray}
To be precise, for any given {\em fixed} time interval $T$,
$T\hbar\dot{\varphi}\sim 2\pi\hbar$
which is invariant under the uniform scale transformation 
$y_{k}(t)\rightarrow 
\epsilon y_{k}(t)$. On the other hand, 
one has $T gr\sin\theta \rightarrow T\epsilon gr\sin\theta$
by the above scaling, and thus one can choose 
$T\epsilon gr\ll \hbar$.

In this new basis (3.16), the geometric phase appears only for
 the mode $c_{+}$ which gives rise to a phase factor
\begin{eqnarray}
\exp\{i\int_{C} \dot{\varphi}dt \}=\exp\{2i\pi \}=1,
\end{eqnarray}
and thus no physical effects. In the infinitesimal neighborhood 
of level crossing, the states spanned by 
$(b_{+},b_{-})$ are transformed to a linear combination of the 
states spanned by $(c_{+},c_{-})$ as is specified by (3.13), which give no non-trivial geometric phases. We thus 
conclude~\cite{fujikawa, fujikawa2} that
 {\em geometric phases are  topologically trivial} for any fixed finite $T$.
The transformation from $b_{\pm}$ to 
$c_{\pm}$ is highly non-perturbative in the sense of adiabatic 
approximation, since a complete rearrangement of two levels is involved.
Incidentally, our analysis shows that the integrability of Schr\"{o}dinger
equation is consistent with the appearance of the seemingly 
non-integrable phases.

It is noted that one cannot simultaneously diagonalize the 
conventional energy eigenvalues and the induced geometric terms in (3.8). The topological considerations are thus inevitably 
approximate. In this respect, it may be instructive to 
consider a model without level crossing which is defined by 
setting 
\begin{eqnarray}
y_{3}=\Delta E/2g
\end{eqnarray}
in
\begin{eqnarray}
h(X(t))
&&=\left(\begin{array}{cc}
            E(0)+y_{0}(t)&0\\
            0&E(0)+y_{0}(t)
            \end{array}\right)
+ g \sigma^{l}y_{l}(t)
\end{eqnarray}
where $\Delta E$ stands for the minimum of the level spacing. 
The geometric terms then loose invariance under the uniform 
scaling of $y_{1}$ and $y_{2}$.
In the limit 
\begin{eqnarray}
\sqrt{y^{2}_{1}+y^{2}_{2}}\gg\Delta E/2g,
\end{eqnarray}
(and thus $\theta\rightarrow \pi/2$), the geometric terms
exhibit approximately topological behavior for the reduced 
variables $(y_{1},y_{2})$. Near the point where the level 
spacing becomes minimum, which is specified by 
\begin{eqnarray}
(y_{1},y_{2}) \rightarrow (0,0)
\end{eqnarray}
(and thus $\theta\rightarrow0$), one can confirm that the 
geometric terms give  trivial phase just as in (3.16).

 Our analysis shows that the model {\em with} 
level crossing exhibits precisely the same topological 
properties for  any finite $T$.  An intuitive picture behind our analysis is that the motion in  $\dot{\varphi}$ smears the 
``monopole'' singularity for arbitrarily large but finite $T$. 

\section{ Concrete example}

It is instructive to analyze a concrete example~\cite{geller, bhandari}  where 
\begin{eqnarray}
(y_{1},y_{2},y_{3})=(B_{0}\cos\omega t, 
B_{0}\sin\omega t, B_{z})
\end{eqnarray}
and $g=\mu$ in
\begin{eqnarray}
h(X(t))
&&=\left(\begin{array}{cc}
            E(0)+y_{0}(t)&0\\
            0&E(0)+y_{0}(t)
            \end{array}\right)
+ g \sigma^{l}y_{l}(t)
\end{eqnarray}
where $B_{0}, B_{z}$ and $\mu$ are constants.
The case $B_{z}\neq 0$ corresponds to
the model without level crossing discussed above, and the 
geometric phase becomes trivial for $B_{0}\rightarrow 0$. 
\begin{figure}[!htb]
 \begin{center}
    \includegraphics[width=10.9 cm]{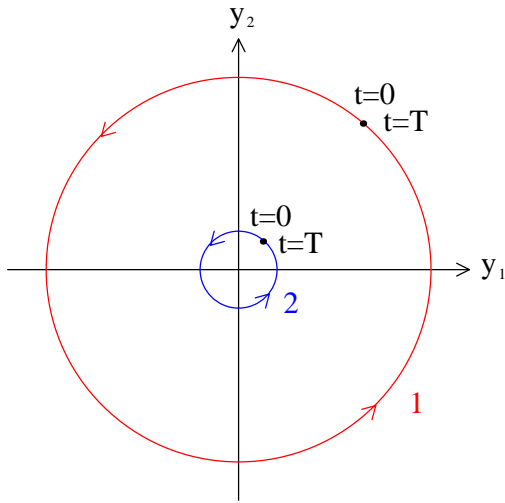} 
       \end{center}
       
\vspace{-9mm}      
 \caption{\small
 The path 1 for $(y_{1},y_{2},y_{3})=(B_{0}\cos\omega t, 
B_{0}\sin\omega t, 0)$ 
gives rise to the phase change rule (namely, $\cos\theta=0$ in
(3.12)) for a fixed finite $T=2\pi/\omega$ 
and  $\mu B_{0}/\hbar\omega\gg 1$, whereas the path 2 gives a trivial 
phase for a fixed finite $T$ and $\mu B_{0}/\hbar\omega\ll 1$, thus 
resulting in the failure of the topological argument for the phase change rule for 
any fixed finite $T$.} 
\end{figure}
\normalsize

The case $B_{z} = 0$ and $B_{0}\rightarrow 0$ describes a cyclic evolution in the infinitesimal neighborhood of level 
crossing, and the geometric phase becomes trivial if 
$T=2\pi/\omega$ is kept fixed. On the other hand, the usual adiabatic approximation (with $\theta=\pi/2$ in the present model) in the neighborhood of 
level crossing is described by $B_{z} = 0$ and 
$B_{0}\rightarrow 0$ (and
$\omega\rightarrow 0$) with 
\begin{eqnarray}
\mu B_{0}/\hbar\omega\gg 1 
\end{eqnarray}
kept fixed, namely, the "effective magnetic field" is always 
strong; the topological proof of 
phase-change rule~\cite{stone} (namely, $\cos\theta=0$ in
(3.12)) is based on the consideration of 
this case. It should be noted that the geometric phase 
becomes trivial for $B_{0}\rightarrow 0$ with $B_{z} = 0$ 
and 
\begin{eqnarray}
\mu B_{0}/\hbar\omega\ll 1 
\end{eqnarray}
kept fixed. (If one starts with $B_{z} = 0$ and 
$\omega=0$, of course, no geometric terms.) 
It is clear that the topology is non-trivial only for a quite 
narrow (essentially measure zero) window of the parameter space
 $(B_{0}, \omega)$ in the approach to the level crossing
$B_{0}\rightarrow 0$. 
\begin{figure}[!htb]
 \begin{center}
    \includegraphics[width=10.9 cm]{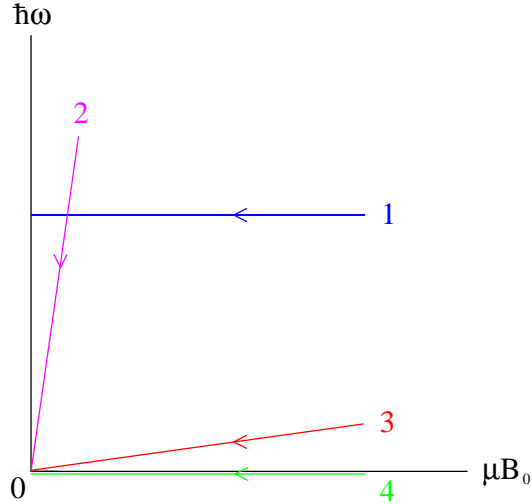} 
       \end{center}
       
\vspace{-10mm}      
 \caption{ 
\small
 Summary of the behavior of the geometric phases in the 
approach to the level crossing point $B_{0}\rightarrow 0$ in the 
parameter space $(\mu B_{0},\hbar\omega)$. The path 1 with fixed 
$\omega=2\pi/T\neq 0$ and also the path 2 with fixed 
$\mu B_{0}/\hbar\omega\ll 1$ give a trivial phase for $B_{0}\rightarrow 0$. 
The path 3 with fixed 
$\mu B_{0}/\hbar\omega\gg 1$ gives a non-trivial phase for 
$B_{0}\rightarrow 0$. The path 4 with $\omega=0$ gives no
geometric phase. The non-trivial phase arises for an essentially
measure zero set in the parameter space 
$(\mu B_{0},\hbar\omega)$ 
for the approach to the level crossing $B_{0}\rightarrow 0$. } 
\end{figure}
\normalsize

The path integral, where the 
Hamiltonian is diagonalized both at $t=0$ and $t=T$ if 
$X(T)=X(0)$, 
\begin{eqnarray}
Z&=&\int \prod_{n}{\cal D}a_{n}^{\star}
{\cal D}a_{n}
\exp\{\frac{i}{\hbar}\int_{0}^{T}dt[
\sum_{n}a_{n}^{\star}(t)i\hbar\frac{\partial}{\partial t}
a_{n}(t)\nonumber\\
&&-\sum_{n,m}a_{n}^{\star}(t)E_{nm}(X(t))a_{m}(t)] \}
\end{eqnarray}
shows no obvious singular behavior at the level 
crossing point.
On the other hand, the path integral with  
$\{v_{n}(\vec{x},X(t))\}$ in (2.8) is  subtle at the 
level crossing point; the bases $\{v_{n}(\vec{x},X(t))\}$ are 
singular on top of level crossing as is seen in (3.5), and thus the induced geometric terms become singular in 
\begin{eqnarray}
&&Z=\int \prod_{n}{\cal D}b_{n}^{\star}{\cal D}b_{n}
\exp\{\frac{i}{\hbar}\int_{0}^{T}dt[
\sum_{n}b_{n}^{\star}(t)i\hbar\frac{\partial}{\partial t}
b_{n}(t)\nonumber\\
&&+\sum_{n,m}b_{n}^{\star}(t)
\langle n|i\hbar\frac{\partial}{\partial t}|m\rangle
b_{m}(t)
-\sum_{n}b_{n}^{\star}(t){\cal E}_{n}(X(t))b_{n}(t)] \}.
\end{eqnarray}

The present analysis, however,  shows that 
 the path integral is not singular for any finite $T$ if one chooses a suitable regular basis near the level crossing point. 
We consider that this result is natural since
 the starting Hamiltonian (3.1) does not contain any obvious 
singularity.

\section{ Discussion}

The notion of Berry's phase is known to be useful in various 
physical contexts~\cite{shapere, review}, and the 
topological considerations are often crucial to obtain a 
qualitative understanding of what is going on. Our 
analysis~\cite{fujikawa, fujikawa2} 
however shows that the geometric phase associated with level 
crossing becomes topologically trivial  in practical physical 
settings with any finite $T$. This is in sharp contrast to the 
Aharonov-Bohm phase which is induced by the 
time-independent gauge potential and topologically exact for 
any finite time interval $T$.
The fact that the geometric phase becomes
topologically trivial for practical physical settings 
with any fixed finite $T$, such as in the practical 
Born-Oppenheimer approximation where a large but finite ratio of two time scales is involved and $T$ is identified with the 
period of the slower system, has not been clearly stated in the 
literature. We emphasize that this fact is proved independently 
of the adiabatic approximation.

Our analysis shows that the notion of the geometric phase is useful, but great care needs to be exercised as to its topological properties. From the present point of view, the essence of geometric phases 
is contained in (2.12): If one evaluates the right-hand side
one obtains an exact result, but the physical picture of what
is going on is not clear. On the other hand, if one makes an approximation (adiabatic approximation) on the left-hand side
of (2.12), one obtains a clear physical picture though the 
result becomes inevitably approximate. 

As for the technical aspect of the present approach, it has been recently emphasized~\cite{fujikawa3} that the notion of hidden local gauge symmetry plays a central role in the actual analyses instead of 
the notions of parallel transport and holonomy.

\end{document}